\documentclass[aps,twocolumn,prd,reprint,nofootinbib,floatfix]{revtex4-2}
\RequirePackage[l2tabu, orthodox]{nag}

\usepackage[utf8x]{inputenc}
\usepackage{adjustbox}

\usepackage{microtype} 
\usepackage{tikz-feynman}
\usepackage{import}
\usepackage{pgfplots}
\usepackage{amsmath}
\usepackage{amstext, amssymb, amsthm, amsfonts, slashed}
\usepackage{physics}
\usepackage{mathtools}
\usepackage{mhchem}
\usepackage{graphicx}
\usepackage{booktabs}
\usepackage{dcolumn}
\usepackage{bm}
\usepackage{dsfont}
\usepackage[normalem]{ulem}
\usepackage{footmisc}

\usepackage{url}
\usepackage[colorlinks=true,urlcolor=blue,linkcolor=blue,citecolor=blue,hypertexnames]{hyperref}
\usepackage[nameinlink]{cleveref}
\makeatletter
\AddToHook{cmd/appendix/before}{%
	\crefalias{section}{appendix}%
	\crefalias{subsection}{appendix}%
}
\makeatother
\usepackage{braket}
\usepackage{simplewick}
\usepackage{float}
\usepackage{subfigure}
\usepackage{dsfont}
\usepackage{comment}

\usepackage{tikz}

\hypersetup{
	colorlinks,
	linkcolor={blue!75!black},
	citecolor={blue!75!black},
	urlcolor={blue!75!black}
}

\crefname{section}{Sec.\!}{Secs.\!}
\crefname{figure}{Fig.\!}{Figs.\!}
\crefname{equation}{}{}
\crefname{table}{Tab.\!}{Tabs.\!}
\crefname{appendix}{App.\!}{Apps.\!}

\usepackage{xcolor,colortbl}

\newcommand{\orcid}[1]{\href{https://orcid.org/#1}{\includegraphics[height=1.7ex,width=1.7ex]{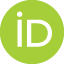}}}

\graphicspath{{figures/}, {}}

\allowdisplaybreaks

\begin{document}

\title{Matter Spectral Functions from Quantum Gravity}

\author{Varun Kher~\orcid{0009-0007-0882-1195}}
\thanks{\href{mailto:v.kher@sussex.ac.uk}{v.kher@sussex.ac.uk}}
\affiliation{Department  of  Physics  and  Astronomy,  University  of  Sussex,  Brighton,  BN1  9QH,  U.K.}

\author{Brandon King~\orcid{0009-0007-6929-4392}}
\thanks{\href{mailto:bk287@sussex.ac.uk}{bk287@sussex.ac.uk}}
\affiliation{Department  of  Physics  and  Astronomy,  University  of  Sussex,  Brighton,  BN1  9QH,  U.K.}

\author{Daniel~F.~Litim\,\orcid{0000-0001-9963-5345}}
\thanks{\href{mailto:d.litim@sussex.ac.uk}{d.litim@sussex.ac.uk}}
\affiliation{Department  of  Physics  and  Astronomy,  University  of  Sussex,  Brighton,  BN1  9QH,  U.K.}

\author{Manuel~Reichert~\orcid{0000-0003-0736-5726}}
\thanks{\href{mailto:m.reichert@sussex.ac.uk}{m.reichert@sussex.ac.uk}} 
\affiliation{Department  of  Physics  and  Astronomy,  University  of  Sussex,  Brighton,  BN1  9QH,  U.K.}

\begin{abstract}
	We investigate  Lorentzian quantum gravity coupled to a template matter sector with gauge fields, scalars and fermions. In the absence of quantised gravity, the matter sector by itself is renormalisable, but UV-incomplete. Provided quantum gravity offers an asymptotically safe UV completion, we determine the photon and scalar two-point functions in the presence of gravitational fluctuations, and show that both possess a K\"all\'en-Lehmann spectral representation. Our results are achieved using functional renormalisation adapted for theories in Lorentzian signature.  We explain why and how interactions with gravity modify both the infrared as well as the ultraviolet behaviour of matter spectral functions. We further determine the corresponding form factors on the level of the quantum effective action. Limitations and extensions of our study are discussed alongside implications for particle physics and unitarity of quantum gravity with matter.
\end{abstract}

\maketitle

\section{Introduction}

The unification of general relativity with quantum mechanics continues to offer challenges. An important contender for a consistent quantum theory of gravity is asymptotically safe gravity \cite{Weinberg:1980gg, Reuter:1996cp}, which retains the metric field as the fundamental carrier of the gravitational force. Then, both gravity and particle physics are fundamentally governed by the laws of quantum field theory, and controlled by a fixed point at the highest energies.

In the past decades, significant evidences for asymptotic safety of gravity have been accumulated starting from quantum Einstein gravity \cite{Souma:1999at, Reuter:2001ag, Litim:2003vp, Fischer:2006fz, Manrique:2011jc, Donkin:2012ud, Falls:2017cze, Baldazzi:2021orb, Kluth:2024lar}, higher-order curvature extensions  \cite{Codello:2007bd, Machado:2007ea, Niedermaier:2009zz, Falls:2013bv, Falls:2014tra, Gies:2016con, Falls:2016wsa, Falls:2017lst, Christiansen:2017bsy, Falls:2018ylp, Kluth:2020bdv, Falls:2020qhj, Knorr:2021slg, Morris:2022btf, Kluth:2022vnq, Baldazzi:2023pep, Falls:2024noj}, graviton vertex functions~\cite{Christiansen:2012rx, Christiansen:2014raa, Christiansen:2015rva, Denz:2016qks, Knorr:2021niv, Bonanno:2021squ, Fehre:2021eob}, and more \cite{Litim:2008tt, Hindmarsh:2011hx, Bonanno:2020bil, Pereira:2019dbn, Reuter:2019byg, Reichert:2020mja, Platania:2020lqb, Pawlowski:2020qer, Knorr:2022dsx, Eichhorn:2022gku, Platania:2023srt, Pawlowski:2023gym}. Equally important, a range of steps have been taken to understand the interplay of quantum gravity with matter \cite{Dona:2013qba, Meibohm:2015twa, Eichhorn:2017sok, Eichhorn:2018akn, Eichhorn:2018ydy, Eichhorn:2018nda,Folkerts:2011jz, Eichhorn:2016esv, Eichhorn:2017lry, Christiansen:2017gtg, Christiansen:2017cxa, Eichhorn:2017eht, Pawlowski:2018ixd, Burger:2019upn}, including predictions of Standard Model (SM) parameters \cite{Shaposhnikov:2009pv, Eichhorn:2017ylw, Eichhorn:2018whv}, UV-complete trajectories \cite{Pastor-Gutierrez:2022nki}, and constraints for models beyond \cite{Litim:2007iu, Gerwick:2011jw, Litim:2011cp, Reichert:2019car, Eichhorn:2020kca, Eichhorn:2020sbo, Kowalska:2020zve, deBrito:2023ydd}. 

Another stream of research has been concerned with unitarity, and the interplay of Lorentzian versus Euclidean signature. The determination of timelike correlation functions from their Euclidean counterparts is a difficult task, 
even more so if the metric field is dynamical and 
with a first-principle definition of the Wick rotation lacking \cite{Baldazzi:2018mtl, Baldazzi:2019kim}. Thus far, this has been addressed using foliated backgrounds  \cite{Manrique:2011jc, Rechenberger:2012dt, Biemans:2016rvp, Biemans:2017zca, Knorr:2018fdu, Eichhorn:2019ybe, Knorr:2022mvn,  Saueressig:2023tfy, Korver:2024sam, Saueressig:2025ypi}, spectral reconstructions assuming the existence of a Wick rotation \cite{Bonanno:2021squ}, or directly in Lorentzian signature \cite{Fehre:2021eob}.
Aspects of scattering amplitudes \cite{Draper:2020bop, Knorr:2020bjm, Knorr:2022lzn, Pastor-Gutierrez:2024sbt}, propagator poles \cite{Platania:2020knd, Platania:2022gtt}, positivity bounds \cite{Knorr:2024yiu, Eichhorn:2024wba,Buoninfante:2024yth}, and other approaches  in Lorentzian space-time  \cite{DAngelo:2022vsh, Banerjee:2022xvi, DAngelo:2023tis, DAngelo:2023wje, Banerjee:2024tap, Thiemann:2024vjx, Ferrero:2024rvi, DAngelo:2025yoy} have also been looked into.

In this light, a key challenge towards understanding unitarity of quantum gravity are computations on backgrounds with Lorentzian signature. A welcome step forward has been achieved in \cite{Fehre:2021eob} by combining functional renormalisation with a mass term cutoff  \cite{Litim:1998nf,Litim:2006ag}. A virtue of the setup is that it avoids spurious cuts or poles in correlation functions. The price to pay is that the flow itself requires counterterms that need to be determined separately. A first application has determined the UV-safe Lorentzian graviton propagator \cite{Fehre:2021eob}, also finding that it admits a healthy spectral representation.

As a next natural step, we pick up on \cite{Fehre:2021eob} to investigate real-world quantum gravity coupled to a template matter sector with  $U(1)$ gauge fields, fermions, and uncharged scalars, mimicking ingredients of the Standard Model (SM). In the absence of quantised gravity, the SM is perturbatively renormalisable yet UV-incomplete due to the notorious triviality problem and Landau poles related to elementary scalars and photons \cite{Callaway:1988ya}. It is then crucial to understand whether the matter sector can be rescued by an asymptotically safe UV completion of quantum gravity, and if so, what its ramifications will be.

Concretely, we study how Lorentzian quantum gravity impacts upon matter field propagators by employing a spectral renormalisation group -- i.e.~functional renormalisation equations with a  Callan-Symanzik-type cutoff suitable for the study of spectral functions in Lorentzian signature \cite{Fehre:2021eob}. Together with the spectral function determined in  \cite{Fehre:2021eob}, and mild additional assumptions, we derive and solve flow equations for the photon and scalar propagators in the UV-complete theory \cite{kher-thesis,king-thesis}. We demonstrate that either of these possesses a  K\"all\'en-Lehmann spectral representation dictated by quantum gravity, which in itself achieves a UV fixed point. We then investigate key features of graviton-induced matter spectral functions, and also address their gauge-dependence. In the infrared (IR) limit, we contrast results with findings from effective theory. We further derive the associated form factors in the quantum effective action and discuss implications for the unitarity of scattering amplitudes.

This work is structured as follows.  We recall the basics of quantum gravity coupled to matter (\cref{sec:setup}), followed by a discussion of our main technical tool, the so-called spectral renormalisation group to study $n$-point correlation functions on backgrounds with Lorentzian signature (\cref{sec:spectral_RG}). After detailing the main approximations (\cref{sec:summary-approximation}), we investigate interacting fixed points and UV-IR connecting trajectories (\cref{sec:FP-and-trajectories}). This is followed by a detailed analysis of matter spectral functions together with form factors in the quantum effective action, and a discussion of scattering amplitudes and scattering spectral functions (\cref{sec:spectral-functions}).  We present our conclusions (\cref{sec:conclusions}), and defer technical aspects on gauge fixing (\cref{app:gauge-fixing}) and explicit flow equations (\cref{app:flow-equations}) into appendices.

\section{Photons and scalars in quantum gravity} \label{sec:setup}
The starting point of our computation is the classical action of a gravity-matter system, which is a subsystem of the Standard Model action. Specifically, we include a U(1) gauge field, a minimally coupled uncharged scalar field, as well as a fermionic field that carries a U(1) charge and has a Yukawa interaction with the scalar field. Later, we choose the Yukawa coupling to resemble the top-Yukawa coupling and the U(1) coupling to resemble the hypercharge coupling. However, the main conclusions of our work hold for generic scalar and U(1) fields.

Our classical action is composed of the Einstein-Hilbert action and a matter action,
\begin{align} \label{eq:full-action}
	S=S_{\text{EH}} +S_\text{matter} \,.
\end{align}
The classical  Einstein-Hilbert action reads
\begin{align} \label{eq:EH-action}
	S_{\text{EH}} =   \frac{1}{16\pi G_\text{N}} \int \! \mathrm  d^4x\sqrt{g}\left(2\Lambda-R\right) + S_\text{gf} + S_\text{gh} \,,
\end{align}
with the classical Newton coupling $G_\text{N}$ and the cosmological constant $\Lambda$. The factor $\sqrt{g}$ is an abbreviation for the absolute value of the determinant of the metric tensor, $\sqrt{g} = \sqrt{|\operatorname{det}g_{\mu\nu}|}$. The action has been augmented to include both the gauge-fixing and ghost action. In this work, we use the de Donder gauge-fixing condition in the harmonic-Feynman gauge, see \cref{app:gauge-fixing} for details. The gauge fixing requires a split of the metric field into a background and fluctuation field. Here, we split the metric linearly into a flat Minkowski background $\eta_{\mu\nu}$ and a fluctuation field $h_{\mu\nu}$,
\begin{align}
	g_{\mu\nu} = \eta_{\mu\nu} + \sqrt{G_\text{N}} \, h_{\mu\nu} \,.
\end{align}
The factor of $\sqrt{G_\text{N}}$ ensures that the fluctuation field has mass dimension one, which is standard for a bosonic field. 

The matter action in \cref{eq:full-action} is given by,
\begin{align}\label{eq:matter-action}
	S_\text{matter}&=\int \! \mathrm d^4x \sqrt{g}\Big(\frac12 \partial_\mu \phi \partial^\mu \phi  -y_t  \phi\bar{\psi} \psi -\frac{1}{4}F^{\mu\nu}F_{\mu\nu} \notag \\*
	&\hspace{2cm} +\bar{\psi}(i \slashed{\nabla}-m_\psi)\psi\Big)+ S_\text{gf,U(1)}\,.
\end{align}
It contains standard kinetic terms for the scalar field $\phi$, the fermionic field $\psi$, and the gauge field $A_\mu$ through the field strength tensor $F_{\mu\nu}$. The interactions are given by the Yukawa term with coupling $y_t$ and the spin-covariant derivative $\slashed{\nabla}$, which includes the coupling $g_Y$. We do not include a four-scalar interaction since that term would not directly contribute to the spectral functions.

The slashed spin-covariant derivative acting on a spinor field reads
\begin{align} \label{eq:nabla-slashed}
	\slashed{\nabla} \psi = g_{\mu \nu} \gamma(x)^\mu \nabla^{\nu} \psi = g_{\mu \nu} \gamma(x)^\mu \left( D^\nu + \Gamma(x)^\nu \right) \psi \, ,
\end{align} 
where $\Gamma^\mu$ is the spin connection and $D_\mu = \partial_\mu - i g_Y A_\mu$ is the standard gauge covariant derivative. For the formulation of spinor fields in curved spacetime, the spin-base invariance formalism has been introduced~\cite{Weldon:2000fr, Gies:2013noa, Lippoldt:2015cea}. It is based on the space-time dependence of the Dirac matrices required by the general anticommutation relation $\lbrace \gamma_\mu, \gamma_\nu \rbrace = 2 g_{\mu \nu} $. This space-time dependence determines the spin connection. 

In \cref{eq:matter-action}, we have also included a gauge-fixing term for the U(1) gauge group. Here we work with a standard Landau gauge, see \cref{app:gauge-fixing}. Our results for the spectral functions will be fully independent of this gauge fixing choice.

\section{Renormalisation group} \label{sec:spectral_RG}
In this section, we recall the definition of the K\"all\'en-Lehmann (KL) spectral representation for the graviton and matter propagators, and introduce our main computational tool, i.e.~functional renormalisation group equations for correlation functions in Lorentzian signature \cite{Fehre:2021eob}. A virtue of our setup is that it preserves spectral representations at all RG scales, while also giving access to spectral functions, including at strong coupling.

\subsection{Spectral functions}
In this work, we consider the spectral function of four different field propagators: the graviton $h_{\mu\nu}$, the scalar field $\phi$, the fermion field $\psi$, and the gauge field $A_\mu$. For each field propagator, we define a propagator factor $G_{\Phi}$ that multiplies a tensor structure and is normalised such that it resembles a standard scalar propagator. For example, for the full graviton propagator $\mathcal{G}_{h}$, we define
\begin{align} \label{eq:graviton-prop-tensor-structure}
	\mathcal{G}_{h}^{\mu\nu\rho\sigma} = \sum_{i=1}^5 c_i G_{h}  \mathcal{T}_i^{\mu\nu\rho\sigma} \,,
\end{align}
where the $\mathcal{T}_i$ are the five tensor structures of the graviton propagator. The coefficients $c_i$ are normalised such that classically $G_{h} = 1/(p^2-2 \Lambda)$, which implies $c_\text{tt}= 32 \pi$ for the transverse-traceless tensor structure. We define similar propagator factors for the other field propagators, see \cref{app:gauge-fixing} for more details. On the quantum level, the propagator factors are full functions of the momentum, $G_\Phi \equiv G_\Phi(p)$.

For the full propagator factors $G_\Phi(p)$, we use the KL spectral representation \cite{Kallen:1952zz, Lehmann:1954xi}. It relates the propagator to the spectral function $\rho_\Phi$ via the relation,
\begin{align}\label{KL representation}
	G_{\Phi}(|\Vec{p}|,p_0) = \int_0^\infty \frac{\lambda  \,\mathrm d\lambda}{\pi}\frac{\rho_\Phi(\lambda,|\Vec{p}|)}{\lambda^2 + p_0^2} \,.
\end{align}
Here, $p_0$ and $|\Vec{p}|$ are the temporal and spatial momenta, respectively, while $\lambda$ is the spectral weight. The spectral function itself can be written as,
\begin{align} \label{eq:spectral function from Prop}
	\rho_{\Phi}(\lambda,|\Vec{p}|) = \lim_{\epsilon \rightarrow 0^+}2 \;\text{Im}\; G_{\Phi}\left(p_0=-i(\lambda + i \epsilon),|\Vec{p}|\right).
\end{align}
The plus superscript indicates that we are using the positive sign convention for the choice of branch cut.  The spectral function can be thought of as encoding the entire energy spectrum of the two-point correlation function. Since Lorentz invariance is not violated by our choice of regulator, see next section, we can conveniently compute the spectral function at vanishing spatial momenta, $\vec{p} = 0$. Classically, the spectral function is simply given by a delta peak representing the on-shell mode of the particle. On the quantum level, it additionally contains a multi-particle continuum representing the quantum fluctuations.

\begin{figure*}[tbp]
	\centering
	\includegraphics[width=.75\linewidth]{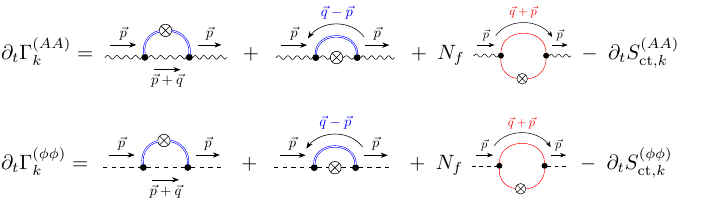}
	\caption{Diagrammatic flow equations for the gauge (top) and scalar field (bottom) two-point function. The double blue lines represent graviton propagators, whereas curly, dashed, or full red lines represent gauge, scalar or fermion propagators, respectively. The number of fermion flavours is indicated by $N_f$. A cross denotes a regulator insertion, and the last terms are flowing counterterm two-point functions \cref{eq:spectral-RG}. Tadpole diagrams, including the graviton-loop tadpole, vanish in our scheme.}
	\label{fig:diagrams}
\end{figure*}

\subsection{Spectral renormalisation group}
Next, we recall the spectral renormalisation group as first laid out in \cite{Fehre:2021eob}, and extend the setup to also include matter spectral functions such as those given in \cref{KL representation}. In its core, our method is a version of the functional renormalisation group (fRG) \cite{Wetterich:1992yh, Morris:1993qb, Ellwanger:1993mw}, which by itself is based on a modified dispersion relation, $p^2 \to p^2 + R_k(p^2)$ due to the Wilsonian regulator function $R_k$. The latter implements the method of integrating out momentum shells around the RG scale $k$. In general, standard Wilsonian regulator functions $R_k(p^2)$ are not compatible with the KL spectral representation \cref{KL representation} since they introduce cuts and poles in the complex momentum plane of the propagator, see e.g.~\cite{Litim:2000ci} for a selection of regulator functions. However, these regulator non-analyticities vanish in the limit $k \to 0$, and it is then possible to perform an analytic continuation \cite{Bonanno:2021squ}. To avoid cuts and poles in the complex plane from the outset, one can either use a regulator that only depends on the spatial momentum and accept the breaking of Lorentz invariance at finite RG scale, or use the Lorentz invariant Callan-Symanzik (CS) regulator \cite{Fehre:2021eob} (see also \cite{Braun:2022mgx}). In a related vein, it has also been noted that a general regulator $R_k(p^2)$ is not compatible with gauge invariance in the sense that it requires the introduction of modified Ward or Slavnov-Taylor identities \cite{Litim:1998nf, Freire:2000bq}, and that it interferes with the structure of thermal fluctuations \cite{Litim:2001up}. Interestingly, either of these aspects are overcome by using a momentum-independent mass term regulator \cite{Litim:1998nf, Litim:2006ag}.

For these reasons, we employ a simple Callan-Symanzik-type regulator in this work. For bosonic fields, it is given by
\begin{align} \label{eq:CS-reg}
	R_{k,\Phi}=Z_{\Phi} \,k^2\,,
\end{align}
where $Z_{\Phi}$ is the on-shell wave function renormalisation of the field $\Phi$. For fermions, the CS regulator is given by
\begin{align} \label{eq:CS-reg-fermion}
	R_{k,\psi}=Z_{\psi}\,k \,\mathds{1} \,,
\end{align}
where $\mathds{1}$ is the identity matrix in Dirac space. This choice of regulator shifts the on-shell contributions, $m_\Phi \rightarrow m_\Phi  + k$, and avoids non-trivial cuts and poles in the complex propagator plane. This makes the CS regulator ideally suited for the computation of spectral functions, which has led to the successful computation of the graviton spectral function \cite{Fehre:2021eob}.

The price to pay for using the CS regulator is that it does not provide a UV regularisation of the flow compared to standard regulator functions \cite{Litim:1998nf}. Hence, standard UV divergences similar to perturbation theory appear and must be treated with counterterms. Currently, there is no known regulator that satisfies causality, Lorentz invariance, and UV finiteness simultaneously \cite{Braun:2022mgx}.

Absorbing the UV divergences in parameters of the scale-dependent effective action $\Gamma_k$ leads to a functional RG equation similar to the standard Wetterich equation \cite{Wetterich:1992yh} but with the inclusion of a running counterterm action $\partial_tS_{\text{ct},k}[\Phi]$. Therefore, the spectral RG equation is given by \cite{Fehre:2021eob, Braun:2022mgx}, 
\begin{align}\label{eq:spectral-RG}
	\partial_t \Gamma_k[\Phi] = \frac{1}{2}\operatorname{Tr}\mathcal{G}_k[\Phi]\;\partial_t R_k - \partial_t S_{\text{ct},k}[\Phi] \,.
\end{align}
Here, the RG time is defined as $t=\ln{k/k_\text{ref}}$ where $k_\text{ref}$ is some reference scale. The full field- and scale-dependent propagator is defined as,
\begin{align}\label{eq: Propagator in terms of effective action}
	\mathcal{G}_k[\Phi]= \frac1{\Gamma_k^{(2)}[\Phi]+R_{k}}\,,
\end{align}
where $\Gamma_k^{(2)}=\delta^2\Gamma_k/\delta\Phi\delta\Phi$. 

We still have a choice on how to regularise the UV divergences. The most convenient choice is dimensional regularisation in $d= 4-2\varepsilon$ spacetime dimensions, which preserves gauge and diffeomorphism invariance. A UV divergence shows up as a $1/\varepsilon$ term, and for each UV divergence, we have to specify a renormalisation condition that also determines the finite part. In this work, we compute the flow of two-point functions, and each of these contains two UV divergences, a quadratic and a logarithmic one. We choose the renormalisation conditions at vanishing momentum, where we fix
\begin{align}  \label{eq:RC}
	\partial_t\Gamma_k^{(\Phi\Phi)}\bigr|_{p=0} &= 0 \,,
	&
	\partial_{p^2} \!\left( \partial_t\Gamma_k^{(\Phi\Phi)}\bigr|_{p=0} \right) &= 0 \,.
\end{align}
Other renormalisation points are possible and, for example, the on-shell renormalisation at $p^2 =-m_\Phi^2$ is a suggestive choice. See \cite{Horak:2020eng, Horak:2021pfr, Horak:2022myj, Braun:2022mgx} for applications of the method beyond gravity.

\subsection{Running spectral functions}
Within this setup, we can derive the flow equation for a spectral function. Taking a derivative of \cref{eq:spectral function from Prop} with respect to the RG time, we get 
\begin{align}\label{eq:flow of the spectral function}
	\partial_t \rho_\Phi(\lambda) = -2\; \text{Im} \, G^2_{\Phi} \left(\partial_t \Gamma_k^{(\Phi\Phi)} + \partial_t R_k\right),
\end{align}
where the right-hand side is evaluated at $p \rightarrow -i(\lambda + i \epsilon)$. 

The flow of the two-point function $\partial_t \Gamma_k^{(\Phi\Phi)}$ can be directly derived from \cref{eq:spectral-RG} by taking two functional derivatives with respect to the field. The corresponding diagrammatic flow equations for the gauge and scalar two-point functions are given in \cref{fig:diagrams}. The first two diagrams show the correction to the propagators via the graviton contributions, and the last diagram represents the correction by fermionic contributions. In principle, the graviton-loop tadpole diagrams (not displayed in \cref{fig:diagrams}) can also contribute terms of the form $\propto \#_1 + \#_2 \, p^2$ to the flow. In practice, however, the tadpole contribution are of the exact same form as those from the $1/\epsilon$ divergences that arise in the renormalisation of the flow. Since our subtraction scheme sets the finite parts of $1/\epsilon$ poles to zero, the contributions from tadpole diagrams are removed by renormalisation through the running counterterm action, see \cref{eq:RC}. This is reminiscent to the vanishing of tadpoles of massless particles in dimensional regularisation.

Importantly, all propagators in \cref{fig:diagrams} are replaced by their respective KL spectral representation. Schematically, the diagrams now have the form
\begin{align}\label{eq: explicit flow equation for two point function}
	\partial_t \Gamma_k^{(\Phi\Phi)} (p) \propto \prod_{i=1}^{3} \int_0^\infty \!\frac{\lambda_i \,\mathrm d \lambda_i}{\pi} \rho_i(\lambda_i) \int \!\frac{\mathrm{d}^d q}{(2\pi)^d} I(p,q,\lambda_i)\,,
\end{align}
where the $\rho_i$ are the spectral functions of the fields that run in the loop, and $I$ is a function that stems from the tensor contractions of the respective diagram and depends on the external momentum $p$, the loop momentum $q$, and the spectral values $\lambda_i$. Together with \cref{eq:flow of the spectral function}, this leads to coupled integro-differential equations for the spectral functions that we solve in this work. 

The diagrammatic flow given in \cref{fig:diagrams} shows that there are two distinct thresholds of the multi-particle continuum of the gauge and scalar spectral function. There is one from the graviton contributions and another from the fermion contributions in the scalar and gauge two-point function. The graviton contribution has a scalar/gauge field and a graviton running in the loop, and therefore the threshold is given by $\theta(\lambda^2-(m_\Phi+m_h)^2)$, while the diagram with the fermion loop has the threshold $\theta(\lambda^2- 4 m_\psi^2)$. Together with the on-shell delta peak, this leads to the following ansatz for the spectral function of the gauge and scalar field,
\begin{align}\label{eq:spectral-function-ansatz}
	\rho_{\Phi} &=  \frac{1}{Z_\Phi}\!\Big[ 2 \pi \delta(\lambda^2-m^2_\Phi) +\theta(\lambda^2- (m_\Phi + m_h)^2)  f_{\Phi,\text{grav}}(\lambda)\notag \\  
	& \hspace*{1.2cm} +\theta(\lambda^2- 4 m_\psi^2)  f_{\Phi,\text{ferm}}(\lambda)\Big],
\end{align}
which holds for both $\Phi =\{A,\phi\}$. Here, $Z_\Phi$ is the wavefunction renormalisation defined at the on-shell point $Z_\Phi(p^2 = -m_\Phi^2)$. Note that the wave function renormalisation factors out in the final flow equation as all propagators, vertices and regulators scale with the respective power of the wave function renormalisations. The dependence on the wave function renormalisation enters through the on-shell anomalous dimension, which is defined by
\begin{align}\label{eq:anom-dim-def}
	\eta_\Phi = -\partial_t  \ln Z_\Phi\,.
\end{align}
The anomalous dimension enters all diagrams through the regulator insertion $\partial_t R_k$.

In \cref{eq:spectral-function-ansatz}, we have displayed scale-dependent mass terms for all fields, including massless fields such as the graviton and the gauge field. The scale-dependent mass terms are made up of two contributions. The first is the direct regulator contribution, see \cref{eq:CS-reg}, and the second one is due to the symmetry breaking of the regularisation and is parameterised with a mass parameter $\mu_\Phi$. In summary, the masses of the bosonic fields are given by
\begin{align} \label{eq:mass-scalar-gauge}
	m_\Phi^2 =k^2(1 + \mu_\Phi(k)) \,.
\end{align}
For the fermionic fields, the masses read
\begin{align}  \label{eq:mass-fermion}
	m_\psi = k(1+\mu_\psi(k))\,.
\end{align} 
In the physical limit, $k\to 0$, the masses of the gauge field and graviton need to vanish, which we can ensure with appropriate boundary conditions for the flows. In comparison, we can choose the physical masses for the scalar and fermion fields.

\section{Summary of approximations} \label{sec:summary-approximation}
In this section, we summarise our setup and list all  couplings and functions, starting with the running couplings and masses
\begin{align} \label{eq:couplings-dimensionful}
	G_\text{N},\, \Lambda, \, y_{t}, \, g_{Y}, \, m_A,\,  m_\phi,\,  m_\psi \,. 
\end{align}
For the analysis of fixed points, we need the corresponding dimensionless versions of the couplings. For the mass parameters of the scalar, fermions, and gauge field, this is given in \cref{eq:mass-scalar-gauge,eq:mass-fermion}. The flow equations for $\mu_\phi$ and $\mu_A$ are extracted by projecting on the $\delta'$ contribution of the respective flow of \cref{eq:flow of the spectral function}. In this work we employ a ``quenched'' approximation, where the feedback of the fermion loop to the flow equations of $\mu_\phi$ and $\mu_A$ is neglected. The resulting equations are presented explicitly in \cref{app:flow-equations}.

We do not compute a flow equation for the fermion mass parameter $\mu_\psi$ since our focus is the computation of scalar and gauge spectral functions, and the leading-order fermion contribution is independent of the details of the fermion mass parameter flow. Therefore, we choose a simple trajectory for it, 
\begin{align}\label{eq:fermion mass parameter}
	\mu_\psi = \frac{c_1 k + c_2 M_\text{Pl}}{k} \,.
\end{align}
This choice implies that the fermion mass parameter is constant above the Planck scale with a fixed-point value $\mu_\psi^* = c_1$, and it is proportional to $k^{-1}$ below the Planck scale. The physical mass of the fermion is given by $c_2 M_\text{Pl}$. In this way, varying the parameters $c_1$ and $c_2$ allows us to check the dependence on the fixed-point value and the physical mass. In our results for the scalar and gauge spectral functions, the fixed-point value $c_1$ has a strongly subleading influence while the physical mass of the fermion is quite important, see \cref{sec:spectral-functions}.

The dimensionless version of the Newton coupling and the cosmological constant are given by 
\begin{align}
	g_{\rm N} &=k^2 G_\text{N} \,, 
	&
	\mu_h &=  - 2\Lambda/k^2 \,.	
\end{align}
In our case, the cosmological constant is related to the dimensionless mass parameter of the fluctuating graviton $\mu_h$. Note that in general there exist several ``avatars'' from which the Newton coupling and the cosmological constant can be extracted, related by Nielsen and Slavnov-Taylor identities, see \cite{Eichhorn:2018akn, Eichhorn:2018ydy, Pawlowski:2020qer}. In this study, we extract the Newton coupling from the three-point vertex and the cosmological constant from the graviton two-point function. The cosmological constant avatars from higher graviton $n$-point functions are set to zero, which was found to be a good approximation in \cite{Christiansen:2015rva, Denz:2016qks}. In our quenched approximation, additionally, the feedback of matter loops on the Newton coupling, the mass parameters, and the anomalous dimensions is neglected. This implies that the flow equations for the Newton coupling and the graviton mass parameter are identical to the results in \cite{Fehre:2021eob}. The full beta functions are presented explicitly in \cref{app:flow-equations}.

The Yukawa and the gauge couplings are essential in our computations as they give the vertex strength within the fermion loops in the scalar and photon field corrections, respectively, see \cref{fig:diagrams}. Since the matter sector serves as a genuine template for models of particle physics, and with a view on the Standard Model, we interpret the Yukawa coupling as a template for the top-Yukawa $y_t$, and the $U(1)$ gauge interaction as a template for the hypercharge coupling $g_Y$. Further, for the numerical analysis below, we use their one-loop beta functions, matched to Standard Model values at the scale $k= 10^{-4} M_\text{Pl}$, see, e.g., \cite{Buttazzo:2013uya}, but neglect the feedback from other matter couplings. This implies a slightly atypical behaviour for the Yukawa coupling around and below the Planck scale compared to the Standard Model, since we neglect the non-abelian gauge contributions that drive the Yukawa coupling to smaller values at higher energies. However, the latter is subleading compared to the dominating graviton contributions above the Planck scale.

\begin{figure*}[tbp]
	\centering
	\includegraphics[width=\linewidth]{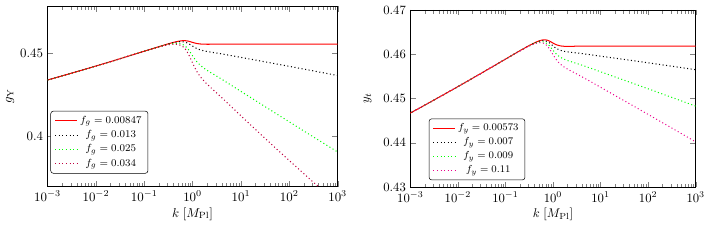}
	\caption{Example trajectories for the gauge (left) and Yukawa coupling (right). The asymptotically safe trajectory is displayed in red together with three asymptotically free example trajectories. All trajectories show a peak-like structure around the Planck scale, which can be attributed to the complex conjugate critical exponent of the Newton coupling.}
	\label{fig:Gauge and Yukawa running}
\end{figure*}

Without gravity, both the U(1) gauge and the Yukawa coupling run into a Landau pole. Quantum gravity needs to provide a UV completion for these couplings. Commonly, the gravity contribution to these couplings is parameterised with coefficients $f_{y}$ and $f_{g}$ \cite{Eichhorn:2017lry} such that the beta functions read 
\begin{align}\label{eq:beta-gauge-Yukawa}
	\beta_{y_t} &= \frac{9}{32\pi^2}y_t^3-f_{y} \, g_{\rm N} \, y_t \,, \notag \\
	\beta_{g_Y} &= \frac{41}{96\pi^2} g_Y^3 - f_{g}\, g_{\rm N}\, g_Y \,,
\end{align}
which we use together with the initial values,
\begin{align} \label{eq:matter-coupling-IR}
	g_Y(k= 10^{-4} M_\text{Pl}) &  = 0.43 \,, \nonumber\\
	y_t(k= 10^{-4} M_\text{Pl} ) & = 0.44  \,.
\end{align}
The coefficients $f_{y}$ and $f_{g}$ depend in principle on gravity couplings such as the graviton mass parameter (or cosmological constant) and higher-order terms in the Newton coupling. At leading order, the coefficients can be approximated as a constant. The combination $f_ig_{\rm N}$ is negligible below the Planck scale, and we have the standard running without gravity, while $f_i g_{\rm N}$ is roughly a constant above the Planck scale. A UV completion of the Yukawa coupling and gauge coupling requires $f_i>0$. Then the beta functions in \cref{eq:beta-gauge-Yukawa} allow for an interacting UV fixed point at a positive coupling value, and the Gau\ss ian fixed point is UV attractive. For $f_i<0$, the  Gau\ss ian fixed point is UV repulsive and the only UV-complete solution is the trivial one, $y_t \equiv 0 \equiv g_Y$.

The coefficients $f_i$ are scheme-dependent since they represent the beta function contribution from a dimensionful coupling, the Newton coupling. It has also been demonstrated that, owing to a kinematic identity, the coefficient $f_g$ is in fact positive or zero $f_g \geq 0$, for any renormalisation scheme \cite{Folkerts:2011jz}, supporting the view that quantum gravity can provide a non-trivial UV completion of the U(1) sector of the Standard Model. Scheme dependence and positivity have been confirmed in a range of model studies, e.g.~\cite{Daum:2009dn, Harst:2011zx, Christiansen:2017gtg, Christiansen:2017cxa, Eichhorn:2017lry, Eichhorn:2019yzm, Pastor-Gutierrez:2022nki, deBrito:2022vbr}, and further identities supporting $f_g \geq 0$ \cite{Folkerts:2011jz} have been found for higher $n$-point functions \cite{Pastor-Gutierrez:2022nki}.

The situation is more intricate in the Yukawa sector. At leading order, $f_y$ is negative at the UV fixed point \cite{Zanusso:2009bs, Oda:2015sma, Eichhorn:2016esv, Eichhorn:2017eht, deBrito:2022vbr} but displays a strong scheme dependence \cite{Eichhorn:2017eht, Pastor-Gutierrez:2022nki}. Only at next-to-leading order, a relevant direction at the Gau\ss ian Yukawa fixed point is found \cite{deBrito:2025nog}. Here, we approximate the next-to-leading order result by using a positive leading-order $f_y$ coefficient. At leading order, a positive value of $f_y$ not only provides a UV completion to the Yukawa sector but could furthermore postdict the mass of the top quark \cite{Eichhorn:2017ylw} or the mass ratio between the top and bottom quark \cite{Eichhorn:2018whv}.

For the scalar sector, we note that it also requires a UV completion by gravity \cite{Shaposhnikov:2009pv}, given that the quartic scalar self-coupling (much like its $U(1)$ and Yukawa counterparts) would otherwise run into a Landau pole. For the sake of this work, we assume that this has been achieved. The specifics of it are irrelevant for the present study, the reason being that quartic self-interactions at the present order of approximation only contribute through tadpoles (see \cref{fig:diagrams}), which are vanishing.

Here, we work in the approximation given in \cref{eq:beta-gauge-Yukawa} and vary the coefficients $f_y$ and $f_g$ such that the couplings become asymptotically safe or free. Corresponding example trajectories are given in \cref{fig:Gauge and Yukawa running}. Within our setup, the UV fixed points of the gauge and Yukawa coupling are given by
\begin{align} \label{eq:matter-coupling-UV}
	g_Y^* &  = 0.455 \,, 
	&
	y_t^* & = 0.462  \,.
\end{align}
The running couplings in \cref{eq:couplings-dimensionful} are supplemented with scale-dependent on-shell wave function renormalisations 
\begin{align} \label{eq:all-Z}
	Z_h, \, Z_\phi, \, Z_A,  \, Z_\psi  \,,
\end{align}
or equivalently, the corresponding anomalous dimensions defined via \cref{eq:anom-dim-def}. Compared to momentum-dependent wave function renormalisations used in, e.g., \cite{Christiansen:2014raa, Christiansen:2015rva, Denz:2016qks}, the wave function renormalisations in \cref{eq:all-Z} are evaluated on-shell, $p^2 = - m_\Phi^2$. The equations for the anomalous dimensions are extracted by projecting on the $\delta$-peak contribution of the respective flow of \cref{eq:flow of the spectral function}. Here, the feedback from the fermion loop is also neglected. The explicit equations are given in \cref{app:flow-equations}. The equation for $\eta_h$ is identical to \cite{Fehre:2021eob}. We neglect the anomalous dimension of the fermion, $\eta_\psi = 0$, which corresponds to $Z_\psi =1$.

\begin{figure*}[tbp]
	\centering
	\includegraphics[width=\linewidth]{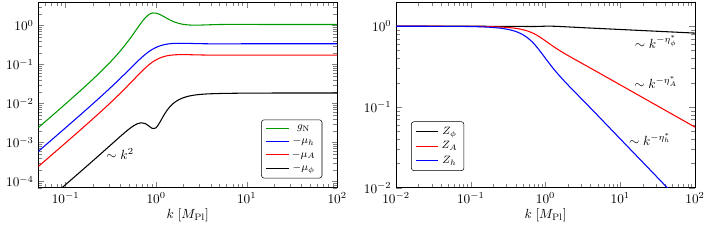}
	\caption{UV to IR connecting trajectories of the Newton coupling and all mass parameters (left) and of the wave function renormalisation (right). The mass parameters and the Newton coupling all scale with $k^{2}$ below the Planck scale and quickly approach their fixed-point values above the Planck scale. The wave function renormalisations are constant below the Planck scale and normalised to 1. Above the Planck scale, they scale with the respective fixed point anomalous dimension $k^{-\eta_\Phi^*}$.}
	\label{fig:UV-IR trajectories-Z and mass}
\end{figure*}

Finally, we compute the running of the multi-particle continua for the scalar and gauge spectral functions
\begin{align} \label{eq:multi-particle-continua}
	f_{\varphi,\text{grav}}, \, f_{\varphi,\text{ferm}}, \, f_{A,\text{grav}}, \, f_{A,\text{ferm}}\,.
\end{align}
Each matter spectral function comes with two distinct threshold functions, one related to the gravity contribution and one related to the fermion contribution, see \cref{eq:spectral-function-ansatz}. In the computation of the flow equations \cref{eq: explicit flow equation for two point function}, we neglect the feedback of the multi-particle continua and only take into account the contribution from the delta peak of the spectral function. This approximation has also been used in \cite{Fehre:2021eob}, and it has the technical advantage that it turns the integro-differential equation into a simple differential equation. We can quantify the quality of this approximation in the deep IR, where we compare to exact one-loop results from effective field theory, see \cref{sec:gravity contributions}. On this quantity, the approximation only introduces an error of 8.4\%.

In summary, the main approximations of this work are the following: 
\begin{itemize}
	\item We use simple flows for the fermion mass $ \mu_\psi $, see \cref{eq:fermion mass parameter}, the hypercharge coupling $g_{Y}$, and the top Yukawa coupling $y_{t}$, see \cref{eq:beta-gauge-Yukawa}. They correctly capture their measured values in the IR, and the cross-over into the asymptotically safe UV fixed point.
	\item We use the ``quenched'' approximation whereupon closed matter loops are neglected. This approximation has been found to work well for few matter fields, up to including Standard Model field contents \cite{Meibohm:2015twa, Christiansen:2017cxa, Eichhorn:2018ydy, Eichhorn:2022gku}.
	\item We neglect the feedback of multi-particle continua $f$ into the flow of spectral functions. This approximation was found to be quantitatively reliable in quantum gravity without matter \cite{Pawlowski:2025etp, Assant:2026dca}. 
\end{itemize}
With these approximations, we determine the flow equations and trajectories for
\begin{subequations} \label{eq:couplings-all}
	\begin{align}
		g_{\rm N},\, \mu_h ,\,  \mu_\phi, \,  \mu_A,  \,  Z_h, \,  Z_\phi,  \, Z_A\,,
	\end{align}
	together with the running multi-particle continua
	\begin{align}
		f_{\varphi,\text{grav}},\,  f_{\varphi,\text{ferm}}, \, f_{A,\text{grav}}, \, f_{A,\text{ferm}}\,.
	\end{align}
	These flow equations were computed with Mathematica and the packages xAct \cite{Martin-Garcia:2008ysv, Brizuela:2008ra} and  HypExp \cite{Huber:2005yg}. All flow equations are displayed in a supplementary Mathematica notebook.
\end{subequations}

\section{Fixed points and Trajectories} \label{sec:FP-and-trajectories}
In this section, we provide results for fixed points, critical exponents, anomalous dimensions, and trajectories connecting the fixed point in the UV with classical general relativity in the IR. We also compare our findings in Lorentzian signature with earlier ones based on Euclidean computations.

The flow equations for couplings \cref{eq:couplings-all} are given in \cref{app:flow-equations} as well as in a supplementary Mathematica notebook. Under the assumption that the Newton coupling is positive and that the mass parameters remain small, we find a unique, interacting UV fixed point at 
\begin{align}
	(g^*_{\rm N}, \, \mu_h^*,\, \mu_A^*, \, \mu_\phi^* ) = (1.06,\, -0.34,\, -0.17,\, -0.018)\,.
	\label{eq:FP-val}
\end{align}
The coordinate for the Newton coupling and the graviton mass parameter agree with \cite{Fehre:2021eob} by construction. What's new is that the extended theory with matter continues to achieve a UV fixed point. Quantitatively, we find that the dimensionless mass parameters $\mu_A^*$ and $\mu_\phi^*$ of the gauge and scalar field both take small, negative values.

For the field anomalous dimensions, we find
\begin{align}\label{eq:FP-anom-dim}
	(\eta_h^*, \eta_A^*,\eta_\phi^*)= (0.96,\, 0.52,\,0.045)\,.
\end{align}
All anomalous dimensions are positive and satisfy the bound $\eta_\Phi < 2$ \cite{Meibohm:2015twa}. We observe that the scalar anomalous dimension comes out an order of magnitude smaller than those of the photon or the graviton. The smallness of scalar field anomalous dimensions has also been observed in Euclidean quantum gravity \cite{Meibohm:2015twa}.

Universal critical exponents are defined as minus the eigenvalues of the stability matrix
\begin{align}
	M_{ij} = \frac{\partial \beta_{\xi_i}}{\partial_{\xi_j}} \,,
\end{align}
evaluated at the fixed point, where $\xi_i$ are all couplings and mass parameters. In our setup, the critical exponents of gravity and matter couplings disentangle, meaning that the gravitational ones
\begin{align}\label{eq:crit-exp-gravity}
	\theta_{1,2} = 2.49 \pm 3.17\, i 
\end{align}
agree with those found in \cite{Fehre:2021eob}. They form a complex conjugate pair of eigenvalues, indicating that vacuum energy and scalar Ricci curvature are strongly correlated operators in the UV.

\begin{figure*}[tbp]
	\centering
	\includegraphics[width=\linewidth]{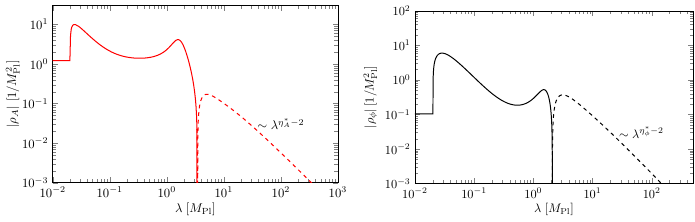}
	\caption{The spectral function of the gauge (left) and scalar field (right) including all diagrams for $N_f = 1$ fermions with a mass of $m_\psi = 10^{-2} M_\text{Pl}$ on the asymptotically safe trajectory of the Yukawa and gauge coupling. The dashed lines indicate that the spectral function is negative. An unrealistically large fermion mass was chosen for the illustration of the fermion threshold behaviour. Both spectral functions are structurally similar and display negative parts in the UV.}
	\label{fig:full-scalar-photon-spectral}
\end{figure*}

\begin{figure*}[tbp]
	\centering
	\includegraphics[width=\linewidth]{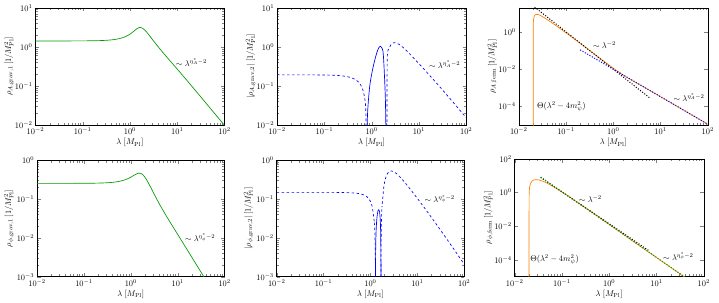}
	\caption{Gauge (top) and scalar (bottom) spectral function for each contributing diagram in the same ordering as in \cref{fig:diagrams}. The individual contributions are structurally very similar, and only the gravity diagram with matter regulator insertion $\rho_{\Phi,\text{grav},2}$ gives negative contributions, which is indicated by the dashed lines. The contribution $\rho_{\Phi,\text{ferm}}$ is universal and dominates below the Planck scale.}
	\label{fig:spectral-individual}
\end{figure*}

In the matter sector, the scaling dimensions of mass parameters are found to be
\begin{align}\label{eq:crit-exp-matter}
	(\theta_{\mu_A}\,, \theta_{\mu_\phi}) = (2.55,\, 2.15)\,.
\end{align}
In our setup, they decouple from each other. Therefore, the deviation from their canonical value (which is two) is entirely due to quantum gravity. The corrections are below 10\% for the scalar field and 30\% for the gauge field -- a comparatively small effect numerically.

We now compute UV-IR trajectories emerging from the fixed point given in \cref{eq:FP-val}. We have four relevant directions at this fixed point. The two relevant directions in \cref{eq:crit-exp-gravity} allow us to fix the physical Newton coupling to its correct IR value and choose any physical cosmological constant. In the present work, we choose $\Lambda =0$. The two relevant directions in \cref{eq:crit-exp-matter} allow us to set the gauge field mass to zero, and choose a physical scalar mass $\omega_\phi$, which we take to be $\omega_\phi=0$. 
In summary, we impose the boundary conditions,
\begin{align}     \label{eq:IR-boundary}
	&\left(G_\text{N}(k),\, k^2 \mu_h(k), \, k^2\mu_A(k), \, k^2\mu_\phi(k) \right)\big|_{k\rightarrow0} \notag \\*[1ex]
	&\hspace{4cm}= (G_\text{N},\, -2 \Lambda, \,  0, \, \omega_\phi)\,,
\end{align}
with $\Lambda = \omega_\phi = 0$. The resulting UV-IR trajectories are displayed in the left panel of \cref{fig:UV-IR trajectories-Z and mass}. 

We solve the on-shell wave function renormalisations on the UV-IR trajectories with the boundary conditions 
\begin{align}    \label{eq:IR-boundary-Z}
	\left(Z_h(k), \,  Z_A(k), \,  Z_\phi(k) \right)\big|_{k\rightarrow0} &= (1, \,  1, \,  1) \,.
\end{align}
The results are shown in the right panel of \cref{fig:UV-IR trajectories-Z and mass}. The wave function renormalisations become constants below the Planck scale, and scale with their anomalous dimension $\propto k^{-\eta_\Phi^*}$ above the Planck scale.

\section{Matter Spectral Functions and Form Factors} \label{sec:spectral-functions}
In this section, we display the gauge and scalar spectral functions, including quantum gravity effects. We provide a brief analysis of their properties and indicate structural similarities between the two matter fields' spectral functions. We then zoom in on their individual diagrammatic contributions and discuss their relation to form factors of the quantum effective action. We provide interpolation functions of the spectral functions in the supplementary Mathematica notebook.

\subsection{Spectral functions}
The spectral functions of the scalar and gauge fields are given by an on-shell delta peak and two multi-particle continua, see \cref{eq:spectral-function-ansatz}. The RG trajectories for the masses and wavefunction renormalisations are given in \cref{fig:UV-IR trajectories-Z and mass}. On these trajectories, we integrate the flow equations for the multi-particle continua \cref{eq:multi-particle-continua}. These flow equations are too lengthy to be displayed but are provided in a supplementary Mathematica notebook. At $k=0$, the on-shell delta peak of the spectral functions is located at vanishing spectral values for the gauge field and at the mass $\omega_\phi$ for the scalar field. We choose the latter to be zero. The multi-particle continuum generated by the gravity diagrams $f_{\Phi,\text{grav}}$ starts at vanishing spectral values since all particles in the loop are massless. The fermion induced multi-particle continuum $f_{\Phi, \text{ferm}}$ starts at $\lambda = 2 m_\psi$.

We display the spectral functions in \cref{fig:full-scalar-photon-spectral} for $N_f=1$ and a fermion mass of $m_\psi = 10^{-2} M_\text{Pl}$. The latter choice is purely for illustrational purposes since the large fermion mass makes it easy to display the threshold effects. This threshold effect is responsible for the peaks around $\lambda = 2\cdot 10^{-2} M_\text{Pl}$. A second peak is visible around the Planck scale and can be traced back to the complex conjugate critical exponents of the Newton coupling \cref{eq:crit-exp-gravity}. This behaviour has also been observed in the graviton spectral function \cite{Bonanno:2021squ, Fehre:2021eob} and in the graviton-mediated $e^+e^- \rightarrow \mu^+\mu^-$ scattering cross section \cite{Pastor-Gutierrez:2024sbt}. For larger spectral values, both matter spectral functions display a change of sign around the Planck scale, followed by an asymptotic scaling $\propto \lambda^{\eta_\Phi^* -2 }$. In general, negative parts may only arise in spectral functions that do not belong to physical observables. Here, the negative parts in  the photon and the scalar field spectral functions reflect their  strong entanglement with the graviton in the deep UV. Therefore, neither of the spectral functions qualify as an observable, much unlike in settings without quantised gravity. In the weak gravity regime, however, the universal fermion loops dominate, spectral functions are positive and remain physical observables for all practical purposes. Further implications and possible remedies of the negative parts are discussed in \cref{sec:form-factors}.

\subsection{Normalisability and gauge-dependence}
Next, we discuss the normalisability and gauge-dependence of spectral functions. The asymptotic scaling $\propto \lambda^{\eta_\Phi^* -2 }$ with $\eta_\Phi^*>0$ implies that spectral functions are decaying more slowly than those with a free propagator $\propto \lambda^{-2 }$, with the immediate consequence that  both spectral functions are non-normalisable,
\begin{align}
	\int_0^\infty \! \lambda \rho_i(\lambda) \, \mathrm d \lambda\,  \neq \text{finite}\,.
\end{align}
A few remarks on the normalisability of spectral functions are in order. Spectral functions that relate to observable asymptotic states need to be positive, and they may or may not be normalisable. Non-normalisable and positive spectral functions have been found for the TT mode of the UV-safe graviton \cite{Fehre:2021eob} using the same techniques as here, and for its reconstructed cousin assuming the validity of a Wick rotation \cite{Bonanno:2021squ}. In models of particle physics (without gravity) with an IR fixed point, positive yet non-normalisable spectral functions have also been found for the gauge and fermion fields  \cite{Kluth:2022wgh}. In our case, the matter spectral functions additionally change sign. Depending on parameters, this scenario can also arise in UV-complete unitary quantum gauge theories that are under strict perturbative control \cite{Kluth:2022wgh}.

We would like to add a few remarks on the properties of the spectral functions. First of all, it is non-trivial that the gauge and scalar propagators possess a KL spectral representation after the inclusion of quantum gravity fluctuations. The propagator could, in principle, show non-analyticities in the complex plane such as complex conjugate poles, which would invalidate the KL representation and be at odds with unitarity, see also \cite{Platania:2022gtt} for a related discussion. We do not observe such behaviour, nor the occurrence of tachyonic or ghost instabilities. 

We further emphasise that the computed spectral functions are not observables and depend on the gauge-fixing parameters of the gravity sector, $\alpha$ and $\beta$. Contrarily, the gauge-field spectral function does not depend on the U(1) gauge-fixing parameter $\chi$ since the U(1) gauge symmetry is linear. As such, the propagators of uncharged scalar fields and the photon, both of which are physical observables in a world without quantised gravity, have become gauge-dependent quantities owing to their cross-talk with quantised gravity.

\begin{figure*}[tbp]
	\centering
	\includegraphics[width=\linewidth]{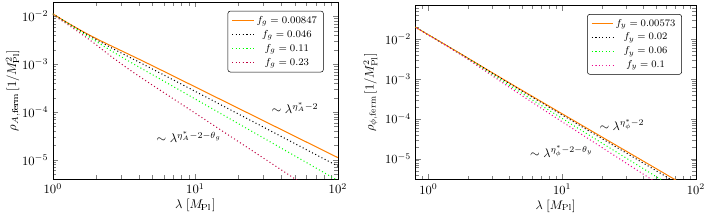}
	\caption{Effect of asymptotically free gauge (left) and Yukawa (right) coupling trajectories on the fermion loop's contribution to matter spectral function. The scaling differs by a factor given by the beta function's critical exponent, which depends on gravity's strength characterised by $f_g$ and $f_y$, see \cref{eq:beta-gauge-Yukawa}. For the values $f_g = 0.00847$ and $f_y = 0.00573$ the respective couplings are asymptotically safe.}
	\label{fig: fermion loop-AF trajectories}
\end{figure*}

The gauge dependence of spectral functions due to the one-loop graviton contribution is already visible in the deep IR limit. They are computed by isolating the $\propto p^4$ contribution to the flow $\partial_t \Gamma_k^{(2)}$ in the weak gravity limit. In perturbation theory, this would correspond to a scheme-independent logarithmic divergence. The same result is obtained within an effective field theory approach to quantum gravity. Still, despite their scheme independence, the one-loop contribution depends on gauge-fixing parameters $\alpha$ and $\beta$ from the gravitational sector. In units of the inverse Planck mass squared, we find
\begin{align} \label{eq:exact-IR}
	\rho_{A,\text{IR}}\big|_{\text{grav}} & =-\frac83 \frac{\beta ^2-3 \alpha}{(\beta -3)^2} \,, \notag \\[1ex]
	\rho_{\phi,\text{IR}}\big|_{\text{grav}} &=6(\beta -1)  \frac{\alpha  (\beta -5)-\beta +9}{(\beta -3)^2} \,.
\end{align}
We recall that the only universal and gauge-independent contribution to spectral functions comes from the fermion loop below the Planck scale. This contribution is positive and dominates between the fermion mass threshold and the Planck scale. However, the one-loop exact IR values \eqref{eq:exact-IR} demonstrate that spectral functions can take positive or negative values in the deep IR, controlled by gauge fixing parameters.  For example, the graviton contribution to the gauge spectral function is the only one remaining below the fermion mass threshold, and it is invariably negative for $\beta> 3\alpha$.

For our choice of gauge-fixing parameters $(\alpha = \beta = 1)$, the one-loop values  \cref{eq:exact-IR} read
\begin{align} \label{eq:exact-IR-11}
	\rho_{A,\text{IR}}\big|_{\rm exact}& = \frac43\,,
	&
	\rho_{\phi,\text{IR}} \big|_{\rm exact}&= 0\,,
\end{align}
which should be compared with our findings from integrating the renormalisation group flow,
\begin{align} \label{eq:our-IR}
	\rho_{A,\text{IR}} \big|_{\rm this\, work}&  \approx 1.22 \,, 
	&
	\rho_{\phi,\text{IR}} \big|_{\rm this\, work}& \approx 0.105  \,.
\end{align}
The difference between \cref{eq:exact-IR-11,eq:our-IR} is due to neglecting the feedback of $f_h,f_A$, and $f_\phi$ into the right-hand side of the flow equation \cref{eq:flow of the spectral function}. Quantitatively, the differences correspond to an absolute shift of $\approx 0.1$, and a relative shift of about $8.4\%$ for the gauge field. The smallness of these differences indicates that having neglected the feedback of $f_h,\,f_A$, and $f_\phi$  in a first instance is a reasonable approximation.

\subsection{Graviton contributions and sign-flip}\label{sec:gravity contributions}
To understand the structure of graviton-induced spectral functions, we disentangle the different diagrammatic contributions. There are two graviton diagrams contributing to the graviton-induced part of the spectral function, see \cref{fig:diagrams}. We show the individual contribution of these diagrams in \cref{fig:spectral-individual}. We observe that the diagram with the graviton regulator insertion is fully positive and dominating in the IR, while the diagram with the matter regulator insertion is mostly negative. The latter diagram dominates in the UV, which explains the sign flip in the full spectral function, see \cref{fig:full-scalar-photon-spectral}. 

Next, we investigate the stability of the sign flip under changes in the underlying approximations. The two contributing diagrams have the UV structure
\begin{align}
	\rho_{\Phi,\text{UV,diag1}} (\lambda) &\propto  \#_1(2- \eta_h^*) \lambda^{\eta_\Phi^* -2} \,,  \notag \\[1ex]
	\rho_{\Phi,\text{UV,diag2}} (\lambda) &\propto  \#_2(2- \eta_\Phi^*) \lambda^{\eta_\Phi^* -2} \,,
\end{align}
where the prefactors $\#_{1,2}$ depend on the Newton coupling and the mass parameters. We now test the stability of the sign change by treating the on-shell graviton anomalous dimension $\eta_h^*$ temporarily as a free parameter. For the gauge and scalar spectral functions to be positive, we find the bounds 
\begin{align}\label{eq:positivity-condition}
	\eta_{h}^*\big|_\text{gauge} &\lesssim 0.3  \,, 
	&
	\eta_h^*\big|_{\text{scalar}} &\lesssim  -54 \,.
\end{align}
As a function of $\eta_h^*$, we note that positivity of the scalar spectral function entails positivity for the gauge spectral function, but not the other way around. Still, the anomalous dimension $\eta^*_h\approx 0.96$ found in  \cref{eq:FP-anom-dim} appears to be too large by a factor of three. In the light of uncertainties due to approximations, however, we conclude that spectral positivity for a UV-complete photon could still be in reach. On the other hand, the non-perturbatively large negative value for $\eta_h^*$ required to achieve positivity for $\rho_\phi$,  \eqref{eq:positivity-condition}, indicates that the sign-change in the scalar sector is hard-wired and spectral positivity out of reach. We stress that extended approximations are not expected to induce quantitatively significant changes in anomalous dimensions. In this light, the conditions  \cref{eq:positivity-condition} can be viewed as estimators for the size of corrections necessary for a {\it qualitative} change in the result.

\begin{figure*}[tbp]
	\centering
	\includegraphics[width=\linewidth]{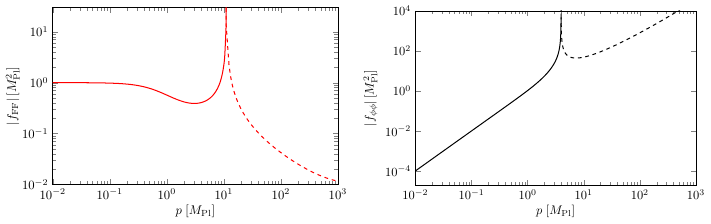}
	\caption{Momentum dependent form factors of the gauge (left) and scalar (right) field with quantum gravity effects computed from their analytical relation with the spectral functions, see \cref{eq: Form factor of photon,eq: Form factor of scalar}. The dashed line indicates negative values.}
	\label{fig: Matter form factors}
\end{figure*}

\subsection{Fermion contributions}
The fermion contribution gives rise to the multi-particle continuum $f_{\Phi, \text{fermi}}(\lambda)$, which starts at the threshold $2m_\psi$. This contribution is universal below the Planck scale, where the gravity contribution to the running coupling can be neglected. We display this part of the spectral function in the right panels of \cref{fig:spectral-individual} for $N_f=1$ computed using the asymptotically safe trajectory for the Yukawa and gauge coupling. In case of multiple fermion flavours, this contribution is simply multiplied by $N_f$. This contribution from the fermion loop is fully positive, and it has two scaling regimes: ${\sim\lambda^{-2}}$ below the Planck scale, and ${\sim\lambda^{\eta^*_\Phi-2}}$ above the Planck scale. In the scalar case, these regimes are difficult to distinguish because the scalar anomalous dimension is small, see \cref{eq:FP-anom-dim}. The precise spectral functions depend on the parameters $c_1,\,c_2$ in the trajectory for the fermion mass parameter, see \cref{eq:fermion mass parameter}. The parameter $c_2$ sets the mass threshold and therefore has a significant impact on the extent of the continuum in the IR. The parameter $c_1$ sets the fixed-point value of the fermion mass parameter, and for reasonable values, the effect is strongly subleading.

The fermion is dominating below the Planck scale due to its scaling behaviour with $\sim \lambda^{-2}$. Above the Planck scale, on the asymptotically safe trajectory, it has the same scaling behaviour as the gravity contribution, but it is subleading compared to it. We want to assess how many fermion flavours are necessary to dominate over the gravity contribution, also in the UV. Structurally, the UV scaling of the diagrams is given by,
\begin{align}
	\rho_{\Phi,\text{UV,grav}} (\lambda) &\propto  \#_1   \,g^*_{\rm N}\, \lambda^{\eta_\Phi^* -2} \,,  \notag \\[1ex]
	\rho_{\Phi,\text{UV,ferm}} (\lambda) &\propto  \#_2\, {\xi^*}^2 N_f \lambda^{\eta_\Phi^* -2} \,,
\end{align}
where $\xi^*$ indicates the fixed-point value of the Yukawa and the gauge coupling, and $\#_{1,2}$ are again some constants. For the fermion diagram to dominate in the UV, we require
\begin{align}
	N_f\frac{({g^*_{Y}})^2}{g^*_N} &\gtrsim \mathcal{O}(100) \,, 
	&
	N_f\frac{({y^*_t})^2}{g^*_N} &\gtrsim \mathcal{O}(1000) \,. 
\end{align}
We therefore conclude that either a very large number of fermion flavours or a highly non-perturbative gauge or Yukawa coupling is needed to make the fermion loop dominate in the UV. Considering that we work in the quenched approximation, our results are certainly not trustworthy for a large number of fermion flavours. Nonetheless, we suspect that the graviton diagrams almost always dominate over the fermion loop ones in the UV. 

So far, we have only been using the asymptotically safe trajectory of the gauge and Yukawa coupling. For the asymptotically free trajectory, the UV scaling of the fermion diagram changes since the diagram is proportional to the respective Yukawa or gauge coupling. The UV scaling for the asymptotically free trajectories is modified by the critical exponent $\theta_{g/y}$ of the gauge/Yukawa beta function \cref{eq:beta-gauge-Yukawa} at their free fixed points respectively. Note that at leading order, the critical exponent is directly given by $\theta_{g/y} =  - f_{g/y}\,g^*_{\rm N}$. The new UV scaling is given by $\lambda^{\eta_A^*-2-\theta_g}$ for the gauge spectral function and $\lambda^{\eta_\phi^*-2-\theta_y}$ for the scalar spectral function. We display the resulting UV behaviour in \cref{fig: fermion loop-AF trajectories}. The result remains identical below the Planck scale, as expected, but shifts above the Planck scale with the exponent $\theta_{g/y}$.

\subsection{Relation to form factors} \label{sec:form-factors}
In the previous sections, we have computed the gauge and scalar spectral functions at the physical scale $k=0$. This information is equivalent to the corresponding two-point functions extracted from the quantum effective action and can be expressed in terms of form factors~\cite{Bosma:2019aiu, Knorr:2019atm, Draper:2020bop, Draper:2020knh, Knorr:2022lzn, Knorr:2022dsx}, see also \cite{Knorr:2021niv, Pawlowski:2023gym, Pawlowski:2023dda, Pastor-Gutierrez:2024sbt}.

In order to highlight the correspondence between form factors and spectral functions, we translate the latter into the former. The part of the quantum effective action that describes the two-point functions is given by \cite{Knorr:2022dsx}
\begin{align} \label{eq:form-factor-action}
	\Gamma = \int \! \mathrm d^4 x \sqrt{g} \left( \frac12 \phi f_{\phi\phi}(\Box)\phi -\frac14 F_{\mu\nu} f_{FF}(\Box) F^{\mu\nu}  \right),
\end{align}
with the two form factors $f_{\phi\phi}$ and $f_{FF}$. Classically, they are given by $f_{\phi\phi} = \Box$ and $f_{FF} = 1$. Including quantum corrections, form factors can be directly extracted from the spectral functions at $k=0$. The scalar form factor is given by
\begin{align}\label{eq: Form factor of scalar}
	f_{\phi\phi} (p^2) = 	G_{\phi}^{-1}(p^2) = \left(\int_0^\infty \!\frac{\lambda  \,\mathrm d\lambda}{\pi}\frac{\rho_\phi(\lambda)}{\lambda^2 + p^2} \right)^{\!-1}\!,
\end{align}
and the photon form factor reads
\begin{align}\label{eq: Form factor of photon}
	f_{FF}(p^2) = 	p^{-2}\, G_{A}^{-1}(p^2) = \left(p^2 \!\int_0^\infty\! \frac{\lambda  \,\mathrm d\lambda}{\pi}\frac{\rho_A(\lambda)}{\lambda^2 + p^2} \right)^{\!-1}\!.
\end{align}
The difference in the equations stems solely from their different definitions in \cref{eq:form-factor-action}, which is also reflected in their classical values.

Due to their simple link to propagators, form factors inherit the very same branch cut structure -- a branch point at $p^2 =0$ and a branch cut along the axis of timelike $p^2$. Moreover, along the axis of spacelike $p^2$ the form factors are real valued. We display the form factors for spacelike momenta in \cref{fig: Matter form factors}. We observe that both the gauge and scalar form factors display a divergence around the Planck scale, becoming negative thereafter. We emphasise that this type of behaviour is unexpected and directly linked to the negative parts in the respective spectral function, see \cref{fig:full-scalar-photon-spectral}. Inasmuch as form factors directly enter the computation of scattering amplitudes, using those found here together with {\it classical} vertices is likely to give unphysical results above the Planck scale, highlighted here for spacelike $t$ or $u$ channel scattering. Hence, our result highlights the necessity to also include quantum gravity corrections for vertices. Provided these are determined in a self-consistent manner, together with corrections to propagators, we may expect that unitarity of quantum gravity allows for cancellations to yield finite and well-defined amplitudes.

The above discussion also highlights the importance of computing the spectral function or form factor related to a scattering amplitude. For example, a photon scattering propagator could be constructed from an electron-positron amplitude,
\begin{align} \label{eq:scat-prop}
	\parbox{1.8cm}{\includegraphics[width=1.8cm]{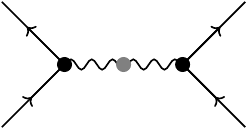}}\Bigg|_\text{amp} &=   Z_A(s) \bar \Gamma^{(\psi\bar\psi A)} \cdot G_A(s) \cdot \bar \Gamma^{(\psi\bar\psi A)} \notag \\*
	&\equiv S^{(\psi\bar\psi A)} \cdot G_{A,\text{scat}}(s) \cdot S^{(\psi\bar\psi A)} \,,
\end{align}
where we have suppressed indices and vertex arguments for easy readability. The last line defines  the ``scattering photon propagator'' $G_{A,\text{scat}}$ as a function of the Mandelstam variable $s$, while the renormalised vertex $\bar \Gamma$ is defined as 
\begin{align}
	\Gamma^{(\psi\bar\psi A)} = \sqrt{Z_A(p_A) Z_\psi(p_\psi)Z_\psi(p_{\bar\psi})}\,\bar\Gamma^{(\psi\bar\psi A)}\,.
\end{align}
Its parameterisation ensures that the wave function renormalisations from the amputated external legs are conveniently cancelled out. The diagram is depicted in the $s$-channel, but the construction is also applicable in the $t$ and $u$ channels due to crossing symmetry.\footnote{A similar construction was used in \cite{Bonanno:2021squ} for the background spectral function, and in \cite{Pawlowski:2023dda} for the gravitational background form factors.}

Thus far, we have computed $G_A(s)$ and $Z_A(s)$ entering  \cref{eq:scat-prop}. To find $G_{A,\text{scat}}(s)$, we also need the vertices $\Gamma^{(\psi\bar\psi A)}$. Crucially, we expect $G_{A,\text{scat}}(s)$ to be gauge-independent given that it is directly related to an observable scattering amplitude. In this light, gauge independence would emerge from combining gauge-dependent vertices with gauge-dependent propagators. For these reasons, we also expect that spectral functions corresponding to amplitudes such as \cref{eq:scat-prop} to be positive semi-definite, and the corresponding form factors to be free of divergences. We look forward to explicit checks of these expectations in future works.

\section{Discussion and conclusions} \label{sec:conclusions}
We have provided inroads into spectral functions and scattering amplitudes of elementary matter fields coupled to quantum gravity. Our study builds upon the recent discovery that the graviton propagator in Lorentzian quantum gravity possesses a healthy spectral representation \`a la K\"all\'en-Lehmann \cite{Fehre:2021eob}. Equally important has been the availability of new renormalisation group methods extended for the theory at hand (\cref{fig:diagrams}), that provide genuine access to propagators and vertex functions on backgrounds with Lorentzian signature (\cref{sec:spectral_RG}) without the need of a Wick rotation \cite{Fehre:2021eob}.

We have put our method to work for a template model of particle physics \cref{eq:matter-action}. Without gravity, the template corresponds to an effective rather than a fundamental theory -- much like the Standard Model itself. Still, spectral functions of photons or uncharged scalars are well-defined physical observables, though only for spectral values sufficiently below a UV cutoff  $\Lambda$ and undefined otherwise.

The primary impact of quantum gravity on \cref{eq:matter-action} is to provide a UV completion with U(1) and Yukawa couplings taking free or interacting fixed points (\cref{fig:Gauge and Yukawa running}). It follows that the UV cutoff can be removed, $1/\Lambda\to 0$, illustrated by RG trajectories connecting the UV fixed point with classical general relativity and standard particle physics in the IR (\cref{fig:UV-IR trajectories-Z and mass}). The matter fields can now be viewed as elementary rather than effective.

Quantum gravity also feeds into matter propagators. Most notably, the latter possess a K\"all\'en-Lehmann spectral representation by themselves  (\cref{fig:full-scalar-photon-spectral}) -- a non-trivial result that cannot be taken as a given in a general theory. Matter spectral functions comprise an on-shell delta-peak and two multi-particle continua,  \cref{eq:spectral-function-ansatz}, one each from graviton and fermion diagrams. Remarkably, our results for gauge and scalar spectral functions share many similarities that can be appreciated by looking into the underlying diagrammatic contributions (\cref{fig:spectral-individual}). The fermion contribution is universal, positive, and dominates below the Planck scale down to the fermion mass threshold. Gravitons contribute with either sign and dominate below the fermion mass threshold, and above the Planck scale where matter spectral functions turn negative. The sign change leads to a divergence in the respective form factors (\cref{fig: Matter form factors}).

It is worth noting that spectral functions of the photon or uncharged scalar fields no longer automatically qualify as physical observables once quantum gravity is present. The reason for this is that graviton terms that feed into matter spectral functions genuinely depend on gauge-fixing parameters. They dominate both in the deep UV, above the Planck scale, and in the deep IR, below fermion mass thresholds  \cref{eq:exact-IR}. Still, below the Planck scale, graviton contributions remain parametrically suppressed $\propto\lambda^2/M_{\rm Pl}^2$ and matter spectral functions are effectively universal, at least for all practical purposes.

Looking forward, we have also discussed how spectral functions translate to form factors in the quantum effective action \cref{eq: Form factor of scalar,eq: Form factor of photon}, and how gauge-independent spectral functions are constructed that encode physical scattering processes \cref{eq:scat-prop}. The latter additionally require quantum-corrected vertices to allow for direct tests of unitarity for quantum gravity with matter. Our work has provided a framework and first key ingredients for this, and we look forward to more extensive studies in the future.

\bigskip

\centerline{\bf Acknowledgements}\smallskip
We thank Gabriel Assant for discussions. This work is supported by the Science and Technology Facilities Council under the Consolidated Grant ST/X000796/1 and the Ernest Rutherford Fellowship ST/Z510282/1. 

\appendix

\section{Gauge fixing and classical propagators} \label{app:gauge-fixing}
The Einstein-Hilbert action in \cref{eq:EH-action} is augmented with a gauge-fixing term given by
\begin{align}
	\label{eq:gf-gravity} 
	S_{\text{gf}}[\bar g, h]=\frac{1}{2 \alpha} \int \!\mathrm{d}^4x
	\sqrt{\bar{g}}\; \bar{g}^{\mu \nu} F_\mu F_\nu \,,
\end{align}
with the gauge-fixing condition $F_\mu$
\begin{align}
	\label{eq:gf-condition-gravity}
	F_\mu[\bar g, h] =
	\bar{\nabla}^\nu h_{\mu \nu} -\frac{1+ \beta}{4} \bar{\nabla}_\mu h^{\nu}_{~\nu} \,.
\end{align}
The respective ghost action reads
\begin{align}
	\label{eq:Sghost-gravity}
	S_{\text{gh}}[\bar g, \phi]=\int \!\mathrm{d}^4x
	\sqrt{\bar{g}}\; \bar c^\mu M_{\mu\nu} c^\nu\,, 
\end{align}
with the Faddeev-Popov operator 
\begin{align}
	\label{eq:OpFP-gravity}
	M_{\mu\nu}= \bar\nabla^\rho\! \left(g_{\mu\nu} \nabla_\rho +g_{\rho\nu} \nabla_\mu\right) -\frac{1+\beta}{2} \bar g^{\sigma\rho} \bar\nabla_\mu g_{\nu\sigma} \nabla_\rho\,.
\end{align}
The gauge-fixing sector enforces the introduction of a background metric $\bar g_{\mu\nu}$, which in our case is given by the flat Minkowski metric, $\bar g_{\mu\nu} = \eta_{\mu\nu}$. We work with the gauge-fixing parameters $\alpha = \beta = 1$. With this setup, the graviton propagator is given by \cref{eq:graviton-prop-tensor-structure}, which also defines the graviton propagator factor $G_h(p)$. For the latter we employ a spectral representation.

The U(1) gauge-fixing action is given by
\begin{align}
	S_{\text{gf, U(1)}} = \frac{1}{2 \chi} \int  \! \mathrm  d^4x \sqrt{\bar{g}} (\bar{D}^\mu A_\mu)\,,
\end{align}
where we work with the Landau gauge condition, $\chi \to  0$. Our results are independent of this gauge-fixing choice. Given this gauge condition, the gauge propagator takes the simple form, 
\begin{align}
	\mathcal{G}^{\mu\nu}_{A} = G_{A} \Pi_\text{T}^{\mu\nu}\,.
\end{align}
where $\Pi_\text{T}^{\mu\nu} = \eta^{\mu\nu}- p^\mu p^\nu/p^2$ is the transverse projection operator, and $G_{A}= 1/p^2$ at the classical level.

The scalar field propagator does not contain any tensor structures and therefore just takes the classical form,
\begin{align}
	\mathcal{G}_\phi= G_\phi = \frac{1}{p^2+\omega_\phi^2}\,,
\end{align}
where in our approximation, we choose $\omega_\phi=0$.

The fermion field propagator carries Dirac indices, and the classical propagator takes the form, 
\begin{align}
	\mathcal{G}_\psi= G_\psi \left(\slashed{p} + i m_\psi \mathds{1}\right)\,,
\end{align}
with the standard propagator factor $G_\psi = 1/(p^2 + m_\psi^2)$.

\begin{widetext}
\section{Flow equations} \label{app:flow-equations}
In this appendix, we provide the explicit flow equations for the mass parameters and anomalous dimensions of the scalar and gauge field, as well as the Newton coupling and the graviton mass parameter and anomalous dimension. Those for the multi-particle continua are too lengthy to be displayed, and we provide them in a supplementary Mathematica notebook. Note that we are working in the quenched approximation, and the effect of the fermion loop is only taken into account for the multi-particle continua. The flow equation for the Newton coupling is given by
\begin{align}
	\partial_t  g_{\rm N} &= (2 + 3 \eta_h)\, g_{\rm N} + \frac{g^2_{\rm N} }{\pi}\! \left(-\frac{47 (6-\eta_h)}{114 (1+\mu_h)^2}+\frac{5 \left(8-\eta_h\right)}{38 (1+\mu_h)^3}+\frac{49 (10-\eta_h)}{570 (1+\mu_h)^4}-\frac{598}{285 (1+\mu_h)^5}-\frac{5}{19 }\right).
\end{align}
This flow equation is derived from the Euclidean flow of the transverse-traceless graviton three-point function at $p=0$ with a Litim-type regulator \cite{Litim:2000ci, Litim:2001up} for the gauge-fixing parameters $\alpha = 0$ and $\beta =1$, see \cite{Christiansen:2015rva, Denz:2016qks}. The flow for the Newton coupling and the graviton mass parameter as well as the graviton anomalous dimension are identical to \cite{Fehre:2021eob}. The flow for the graviton mass parameter reads
\begin{align}
	\partial_t \mu_h & = -2 \mu_h - \eta_h+g_{\rm N}(1+\mu_h)(2 - \eta_h) \frac{5 \left( 5\sqrt{3}\pi - 22 \right) }{18\pi} \notag \\[1ex]
	&\qquad+ \frac{2g_{\rm N}}{3\pi (1+\mu_h)}\left(23+16\mu_h-7\mu_h^2+3\sqrt{\frac{1+\mu_h}{\mu_h-3}}\left(13-6\mu_h+\mu_h^2\right)\text{arcosh}\!\left[\frac{1}{2}\left(1-\mu_h\right)\right]\right),
\end{align}
and the graviton anomalous dimension is given by
\begin{align}
 \eta_h &=  g_{\rm N} (2 -\eta_h) \frac{5 \pi  \sqrt{3}+147}{54\pi}  \notag \\[1ex]
	&\qquad -   \frac{2 g_{\rm N} }{3 \pi  (\mu_h-3) (\mu_h+1)} \left(4 \left(\mu_h^2+\mu_h-15\right) +\frac{3 (\mu_h ((\mu_h-3) \mu_h-13)+31)\, \text{arcosh}\!\left[\frac{1}{2}\left(1-\mu_h\right)\right]}{\sqrt{(\mu_h-3) (\mu_h+1)}} \right). 
\end{align}
The flow equation for the scalar mass parameter is given by
\begin{align}
	\partial_t\mu_\phi&= - 2 \mu_\phi  - \eta_\phi \notag \\
	&\,\quad- 
	\frac{g_{\rm N}  \left( (\eta_\phi -2)(1 + \mu_h)^2 + (8 + \eta_h - 5 \eta_\phi)(1 + \mu_h)(1+\mu_\phi) - 2 (\eta_h-2)(1 + \mu_\phi)^2 \right) 
		\operatorname{artanh}\!\left[ \sqrt{\frac{1 + \mu_h}{-3 + \mu_h - 4 \mu_\phi}} \right]}
	{\pi \sqrt{(1 + \mu_h)(-3 + \mu_h - 4 \mu_\phi)}}\notag \\[10pt]
	&\,\quad - \frac{g_{\rm N}  \left( ( \eta_h - 2) (1 + \mu_h)^2 + (8 + \eta_h - 5 \eta_\phi)(1 + \mu_h)(1+\mu_\phi) - 2 (\eta_h-2)(1 + \mu_\phi)^2 \right)
		\operatorname{artanh}\!\left[ \frac{-1 + \mu_h - 2 \mu_\phi}{\sqrt{(1 + \mu_h)(-3 + \mu_h - 4 \mu_\phi)}} \right]}
	{\pi \sqrt{(1 + \mu_h)(-3 + \mu_h - 4 \mu_\phi)}} \notag\\[10pt]
	&\,\quad- \frac{g_{\rm N} }{ \pi (\mu_h - \mu_\phi)^2}  \Bigg((1+\mu_\phi)(\mu_\phi-\mu_h)((\eta_\phi-2)(1+\mu_h)+(2+\eta_h-2\eta_\phi)(1+\mu_\phi))+\frac{1}{2}\ln( \frac{1+\mu_h}{1+\mu_\phi})\Big((\eta_\phi-2)(1+\mu_h^3)\notag \\
	&\,\qquad+(8+\eta_h-5\eta_\phi)(1+\mu_h)^2(1+\mu_\phi)+(-2-2\eta_h+3\eta_\phi)(1+\mu_h)(1+\mu_\phi)^2+(-4+3\eta_h-\eta_\phi)(1+\mu_\phi)^3\Big) \Bigg)\,.
\end{align}
The flow equation for the gauge mass parameter reads
\begin{align}
	\partial_t\mu_A &= - 2 \mu_A - \eta_A \notag \\
	&\,\quad+ \frac{	g_{\rm N}  (2 - \eta_h)}{2 \pi (1 + \mu_A) (1 + \mu_h)}
	\Bigg(
	\frac{2 (-1 - 2 \mu_A + \mu_h)^3}{\sqrt{\frac{-3 - 4 \mu_A + \mu_h}{1 + \mu_h}}} \left( \operatorname{arcoth}\!\left[ \sqrt{1 - \frac{4 (1 + \mu_A)}{1 + \mu_h}} \right] 
	+ \operatorname{arcoth}\!\left[ \frac{\sqrt{(- 3 - 4 \mu_A + \mu_h) (1 + \mu_h)}}{1 + 2 \mu_A - \mu_h} \right] \right) \nonumber \\
	&\, \qquad + \frac{(1 + \mu_h)}{(\mu_A - \mu_h)^2}  \Bigg( 
	2 (1 + \mu_A) (\mu_A - \mu_h)^3 + 3 (1 + \mu_A)^2 (3 \mu_A^2 + \mu_A (2 - 4 \mu_h) + (-2 + \mu_h) \mu_h) \nonumber \\
	&\, \qquad - \left( 12 (1 + \mu_A)^4 - 16 (1 + \mu_A)^3 (1 + \mu_h) + 15 (1 + \mu_A)^2 (1 + \mu_h)^2 - 6 (1 + \mu_A) (1 + \mu_h)^3 + (1 + \mu_h)^4 \right) \ln( \frac{1 + \mu_A}{1 + \mu_h}) \Bigg) \Bigg) \nonumber \\
	& \,\quad+ \frac{	g_{\rm N}  (2 - \eta_A)}{2 \pi (1 + \mu_A)}
	\Bigg(
	-\frac{2 (1 + \mu_h) (13 + 20 \mu_A^2 - 8 \mu_A (-4 + \mu_h) + (-6 + \mu_h) \mu_h)}{\sqrt{(1 + \mu_h) (-3 - 4 \mu_A + \mu_h)}} \nonumber \\
	& \,\qquad \times\left( \operatorname{artanh}\!\left[ \frac{1 + 2 \mu_A - \mu_h}{\sqrt{(- 3 - 4 \mu_A + \mu_h) (1 + \mu_h)}} \right] 
	+ \operatorname{artanh}\!\left[ \sqrt{\frac{1 + \mu_h}{ -3 - 4 \mu_A + \mu_h}} \right] \right) \notag \\
	& \,\qquad - \frac{1}{(\mu_A - \mu_h)^2} \Bigg( 
	13 (1 + \mu_A)^4 + 2 (1 + \mu_A) (\mu_A - \mu_h)^3 - 20 (1 + \mu_A)^3 (1 + \mu_h) + 7 (1 + \mu_A)^2 (1 + \mu_h)^2\nonumber \\
	&\,\qquad   + \left( 4 (1 + \mu_A)^4 - 14 (1 + \mu_A)^3 (1 + \mu_h) + 23 (1 + \mu_A)^2 (1 + \mu_h)^2 - 8 (1 + \mu_A) (1 + \mu_h)^3 + (1 + \mu_h)^4 \right) 
	\ln( \frac{1 + \mu_h}{1 + \mu_A}) \Bigg) \Bigg)\,.
\end{align}
The scalar anomalous dimension is given by
\begin{align}
	\eta_\phi&= \frac{g_{\rm N} }{\pi (1 + \mu_h) \left(\mu_h^2 + \mu_\phi (3 + 4 \mu_\phi) - \mu_h (3 + 5 \mu_\phi) \right)^2} \notag\\
	&\,\quad\times\Bigg( (2 - \eta_h)\Bigg[ (1 + \mu_\phi)^2 \Bigg( (1 + \mu_h)^4\ln(\frac{1+\mu_h}{1+\mu_\phi}) (-3 + \mu_h - 4 \mu_\phi)^2 \notag\\
	&\,\qquad+ 2 \Bigg( \text{artanh}\! \left[\sqrt{\frac{1 + \mu_h}{-3 + \mu_h - 4 \mu_\phi}}\right] + \text{artanh}\! \left[\frac{1 - \mu_h + 2 \mu_\phi}{\sqrt{(1 + \mu_h) \, (-3 + \mu_h - 4 \mu_\phi)}}\right] \Bigg) \notag \\
	&\,\qquad \times\sqrt{(1 + \mu_h) \, (-3 + \mu_h - 4 \mu_\phi)} \, (\mu_h - \mu_\phi)^2- 3(1 + \mu_h)^2 (1 + \mu_\phi) - 6 \, (1 + \mu_h) (1 + \mu_\phi)^2 + 8 \, (1 + \mu_\phi)^3\Bigg)\Bigg] \notag\\
	&\;\quad + (2 - \eta_\phi) \Bigg[ -3 \, (1 + \mu_h) \, (-3 + \mu_h - 4 \mu_\phi) \, (\mu_h - \mu_\phi) \, (1 + \mu_\phi)^2\notag\\
	&\,\qquad -\Bigg( \text{artanh}\!\left[\sqrt{\frac{1 + \mu_h}{-3 + \mu_h - 4 \mu_\phi}}\right] + \text{artanh}\! \left[\frac{1 - \mu_h + 2 \mu_\phi}{\sqrt{(1 + \mu_h) \, (-3 + \mu_h - 4 \mu_\phi)}}\right] \Bigg)\sqrt{(1 + \mu_h) \, (-3 + \mu_h - 4 \mu_\phi)} \, (\mu_h - \mu_\phi)^2 \notag \\
	&\,\qquad  - \frac{1}{2}\ln(\frac{1+\mu_h}{1+\mu_\phi}) (1 + \mu_h) \, (-3 + \mu_h - 4 \mu_\phi)^2  \left(2 + \mu_h^2 + \mu_\phi^2 + 2 \mu_h \, (2 + \mu_\phi)\right)\Bigg] \Bigg).
\end{align}
Finally, the photon anomalous dimension reads
\begin{align}
	\eta_A &= \frac{g_{\rm N} }{2\pi(1+\mu_A)^3(\mu_A-\mu_h)^2(1+\mu_h)(-3-4\mu_A+\mu_h)}\notag \\
	&\,\quad \times\Bigg((2-\eta_h)\Bigg[-\frac{(1+\mu_A)}{\sqrt{\frac{-3-4\mu_A+\mu_h}{1+\mu_h}}}\Bigg(-\sqrt{1+\mu_h}\left(-3-4\mu_A+\mu_h\right)^{\!\frac{3}{2}}\big(9(1+\mu_A)^4-12(1+\mu_A)^3(1+\mu_h) \notag \\
	&\,\qquad+3(1+\mu_A)^2(1+\mu_h)^2-\ln(\frac{1+\mu_A}{1+\mu_h})\big(3+2\mu_A(3+\mu_A)+2\mu_A\mu_h-\mu_h^2\big)\big(2+\mu_A(4+3\mu_A)-2\mu_A\mu_h+\mu_h^2\big)\big) \notag \\
	&\,\qquad+(\mu_A-\mu_h)^2(-1-2\mu_A+\mu_h)^2\Bigg(-2\,\text{arcoth}\!\left[\sqrt{1-\frac{4(1+\mu_A)}{1+\mu_h}}\right]\bigg(5+2\mu_A^2-(-2+\mu_h)\mu_h+4\mu_A(2+\mu_h)\bigg)\notag \\
	&\,\qquad+2\Bigg(\text{arcoth}\!\left[\frac{\sqrt{(1+\mu_h)(3+4\mu_A-\mu_h)}}{1+2\mu_A-\mu_h}\right]\big(5+2\mu_A(4+\mu_A)+2\mu_h+4\mu_A\mu_h-\mu_h^2\big)\notag \\
	&\,\quad\qquad+(1+\mu_A)(1+2\mu_A-\mu_h)\sqrt{1-\frac{4(1+\mu_A)}{1+\mu_h}}\Bigg)\Bigg)\Bigg)\Bigg]\notag \\
	&\;\quad+(2-\eta_A)  \Bigg[\frac{1+\mu_A}{\sqrt{1+\mu_h}\left(-(3+4\mu_A-\mu_h)\right)^{\!\frac{3}{2}}}\Bigg(-\left(-(3+4\mu_A-\mu_h)(1+\mu_h)\right)^\frac{3}{2}(-3-4\mu_A+\mu_h)\notag \\
	&\,\qquad \times \big(13(1+\mu_A)^4-20(1+\mu_A)^3(1+\mu_h)+7(1+\mu_A)^2(1+\mu_h)^2+\ln(\frac{1+\mu_h}{1+\mu_A})\big(-4(1+\mu_A)^4\notag \\
	&\,\quad \qquad+10(1+\mu_A)^3(1+\mu_h)-3(1+\mu_A)^2(1+\mu_h)^2+4(1+\mu_A)(1+\mu_h)^3-(1+\mu_h)^4\big)\big)\notag \\
	&\,\qquad-(\mu_A-\mu_h)^2\sqrt{-(3+4\mu_A-\mu_h)(1+\mu_h)}\Bigg(-2(1+\mu_A)(3+4\mu_A-\mu_h)(1+\mu_h)\notag \\
	&\,\quad \qquad \times \big(13+20\mu_A^2-8\mu_A(-4+\mu_h)+(-6+\mu_h)\mu_h\big)\notag \\\
	&\,\qquad-2\Bigg(\text{artanh}\!\Bigg[\frac{1+2\mu_A-\mu_h}{\sqrt{-(3+4\mu_A-\mu_h)(1+\mu_h)}}\Bigg]+\text{artanh}\!\Bigg[\sqrt{\frac{1+\mu_h}{-3-4\mu_A+\mu_h}}\Bigg]\Bigg) \notag \\
	&\,\quad \qquad \times \sqrt{-(3+4\mu_A-\mu_h)(1+\mu_h)}\Big(17+16\mu_A^4-8\mu_A^3(-7+\mu_h)+8\mu_A^2\big(11+\mu_h+2\mu_h^2\big)\notag \\
	&\,\qquad+\mu_h(4+\mu_h(-2+(-4+\mu_h)\mu_h))+8\mu_A\big(8+\mu_h\big(2+\mu_h-\mu_h^2\big)\big)\Big)\Bigg)\Bigg)\Bigg]\Bigg).
\end{align}
\end{widetext}

\bibliographystyle{mystyle}
\bibliography{bibliography}

\end{document}